\documentclass[aps,prd,12pt,notitlepage,nofootinbib,tightenlines]{revtex4-1}
\usepackage{amsmath}
\usepackage{bm}
\usepackage{times}
\usepackage{braket}
\usepackage{color}
\usepackage{epsfig}
\usepackage{slashed}
\usepackage{hyperref}
\usepackage{float}
\usepackage{diagbox}
\usepackage{multirow}
\usepackage{amssymb, amsthm, amsmath}
\usepackage{graphicx}
\usepackage{hyperref}
\usepackage{slashed}

\definecolor{Red}{rgb}{1.,0.,0.}

\definecolor{Blue}{rgb}{0.,0.,1.}

\definecolor{nicered}{rgb}{0.7,0.1,0.1}
\definecolor{nicegreen}{rgb}{0.1,0.5,0.1}

\hypersetup{colorlinks,citecolor=nicegreen,linkcolor=nicered}

\begin{document}
\newcommand{\FIXME}[1]{\textcolor{red}{[FIXME: #1]}}
\newcommand{\Qsl}{Q \hspace{-2.0truemm}/}
\newcommand{\epsilonsl}{\epsilon \hspace{-2.0truemm}/}
\newcommand{\Psl}{P \hspace{-2.0truemm}/}
\newcommand{\beq}{\begin{eqnarray}}
\newcommand{\eeq}{\end{eqnarray}}
\newcommand{\non}{\nonumber \\ }
\newcommand{\psl}{ p \hspace{-2.0truemm}/ }
\newcommand{\qsl}{ q \hspace{-2.0truemm}/ }
\newcommand{\epsl}{ \epsilon \hspace{-2.0truemm}/ }
\newcommand{\nsl}{ n \hspace{-2.2truemm}/ }
\newcommand{\vsl}{ v \hspace{-2.2truemm}/ }
\newcommand{\jpsi}{ J/\psi }
\newcommand{\cala}{ {\cal A} }
\newcommand{\calb}{ {\cal B} }


\def \cpc{ { Chin. Phys. C } }
\def \ctp{ { Commun. Theor. Phys. } }
\def \csb{{ Chin. Sci. Bull. } }
\def \sbu{{ Sci. Bull.  } }

\def \epjc{{ Eur. Phys. J. C} }
\def \ijmpa{ { Int. J. Mod. Phys. A } }
\def \jhep{{ JHEP } }
\def \jpg{ { J. Phys. G} }
\def \mpla{ { Mod. Phys. Lett. A } }
\def \npb{ { Nucl. Phys. B} }
\def \plb{ { Phys. Lett. B} }
\def \ppnp{ Prog.Part. $\&$ Nucl. Phys. }
\def \pr{ { Phys. Rep.} }
\def \prd{ { Phys. Rev. D} }
\def \prl{ { Phys. Rev. Lett.}  }
\def \ptp{ { Prog. Theor. Phys. }  }
\def \zpc{ { Z. Phys. C}  }

\def \thl {{\theta_\ell}}
\def \thK {{\theta_{K^\ast}}}
\def \re{\text{Re}}
\def \im{\text{Im}}
\def \eff{{\text{eff}}}
\def\Sin{\text{sin}}
\def\Cos{\text{cos}}

\title{Soft Gluon Resummation in Double Heavy Quarkonium Production at LHC}
\author{Chong-Yang  Lu}  \email{181002013@stu.njnu.edu.cn}
\author{Dan-Dan Shen } \email{181002016@stu.njnu.edu.cn}
\author{Peng Sun } \email{06260@njnu.edu.cn }
\author{Ruilin Zhu } \email{rlzhu@njnu.edu.cn }

\affiliation{ Department of Physics and Institute of Theoretical Physics,
Nanjing Normal University, Nanjing, Jiangsu 210023, People's Republic of China}
	\date{\today}
\begin{abstract}
In this paper, the soft gluon resummation effect in double heavy quarkonium production at
the LHC is studied. By applying the transverse momentum dependent factorization formalism, the
large logarithms introduced by the small total transverse momentum of heavy quarkonium pair final state system,
are resummed to all orders in the expansion of the strong interaction coupling
at the Next-to-Leading Logarithm accuracy.
We also compare our result with the LHC data.
We find that the distribution shape predicted by resummation calculation is consistent with experimental data very well.
Since this process is mainly initiated by gluon fusion, it supplies us an important channel to study 
the gluon parton property in hadrons.   
\end{abstract}

\maketitle

\section{Introduction}\label{sec:1} 
Heavy quark pair production at hadron collision is one of hot topics for high energy physicists all the time.
It is important for us to understand the production mechanism of heavy quarkonia and the basic property of quantum Chromo dynamics(QCD). 
In addition, this process is mainly initiated by gluon fusion, 
and $J/\psi$ is one of the easiest particles to be observe in experiments. 
Thus it supplies us an important channel to study the gluon parton property in hadrons.   
A lot of theoretical works have been done for this process ~\cite{Schafer:2019ynn,Scarpa:2019ucf,He:2019qqr,Pan:2019sxp,Lansberg:2006dh,Brambilla:2010cs,Andronic:2015wma,Qiao:2009kg,Qiao:2002rh,Li:2009ug,Lansberg:2013qka,Lansberg:2020rft},
and this process can be expressed as:
\begin{align}
p + p(\bar{p})  \rightarrow J/\psi(P_1) + J/\psi(P_2).
\end{align}

The LHCb collaboration has measured double $J/\psi$ production cross sections
with an integrated luminosity $37.5$ pb$^{-1}$, at the center-of-mass energy $\sqrt{S}= 7$ TeV
and at the range of $J/\psi$ rapidity $2<y^{J/\psi}<4.5$~\cite{Aaij:2011yc}.
At LHC RUN II they also reported the measurement of transverse momentum
distribution of double $J/\psi$ system, at $\sqrt{S}= 13$ TeV~\cite{Aaij:2016bqq}.
In addition, CMS, ATLAS and D0 collaborations also made relevant measurements for this process~\cite{Aaboud:2016fzt,Abazov:2014qba,CMS:2013pph}.
These experimental measurements supply a lot of important data for us to study the property of heavy quarkonia.

In the studies of heavy quarkonia, nonrelativistic QCD(NRQCD) has become a basic method that deal with the decay or production of heavy quarkonia~\cite{Bodwin:1994jh}.
It has been widely used in heavy quarkonium production~\cite{Butenschoen:2010px,Fan:2009zq,Li:2008ym,Gong:2008sn,He:2007te,Qiao:2012hp,Zhu:2015qoa,Wang:2015bka,Shen:2020woq}.
In NRQCD, by using the factorization method, a process of hadronization can be divided into short-distance perturbative parts and long-distance nonperturbative matrix elements.
For the former, we can use the perturbative QCD to calculate it order by order.
For the later, they are process-independent, which can be extracted from
experimental data or obtained from non-perturbative method. In addition, these matrix elements are organized in terms
of the velocity expansion in the NRQCD framework. A fixed order perturbative calculation
is performed in orders of both the strong coupling constant $\alpha_s$ and the power of the velocity
for the associated matrix elements. In this work, we only consider the leading power contribution in the velocity enpension series,
which comes from the color-singlet matrix element.

In this work, we will fucus on the transverse momentum($q_\perp=P_1^{\perp}+P_2^{\perp}$)
distribution of the heavy quarkonium pair system in the process $p + p(\bar{p})  \rightarrow J/\psi + J/\psi$,
here the transverse momentum $q_\perp$ distributions are mainly determined by the soft gluon radiations,
especially in the region of small $q_\perp$. In order to obtain a reliable prediction of the $q_\perp$ distribution, we must
take into account the soft gluon shower effect. The soft gluon shower effect brings the large Sudakov
logarithms into all orders of the perturbative expansion, and then breaks the validity of the
perturbative expansion. Therefore, we have to perform an all order transverse momentum dependent(TMD) resummation calculation
based on the TMD factorization theorem~\cite{Collins:1984kg,deFlorian:2001zd,Bozzi:2005wk,Berger:2003pd,deFlorian:2011xf,Ji:2004xq},
and resum these large logarithms into a Sudakov factor.

The rest of this article is scheduled in the following. In section II, we present some calculation methods and techniques. In section III, by using the method of resummation, we obtain the transverse momentum $q_\perp$ distribution of the double $J/\psi$ and $\Upsilon$ pair system. In the last section, a summary is given.

\section{METHOD OF CALCULATION}
\begin{figure}[htbp]
	\centering
\includegraphics[width=.65\textwidth]{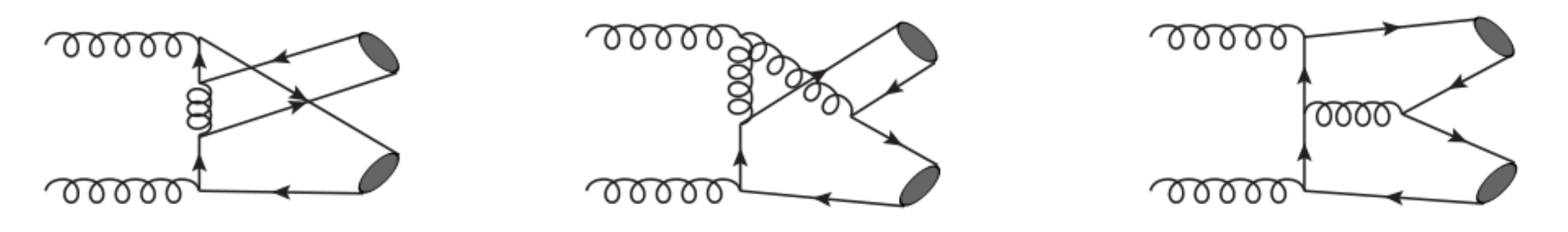}
\caption{The typical Feynman diagrams of double $J/\psi$ in pp collision. }
\label{feyn}
\end{figure}

For double $J/\psi$ production at LHC, the dominant contribution comes from gluon fusion,
the quark anti-quark annihilation processes can be ignored because of the suppression from parton distribution function at high energy scale.
For the NRQCD matrix elements, we only take into account color-singlet operators at the leading order(LO) of the velocity expansion.
The high order color-octet contribution can be ignored, since the main contribution comes kinematical region of each $J/\psi$ with low transverse momentum.
For the process of $g+g\rightarrow{J/\psi}+{J/\psi}$ at parton level, the typical Feynman diagrams of this process are shown in Fig~\ref{feyn}.

The different scattering cross section of $J/\psi$ pair is written as
\begin{align}
&\frac{d\sigma}{dp_T}(pp \rightarrow {2J/\psi})= \sum_{a,b}\int dy_1 dy_2 f_1(x_1) f_2(x_2) 2p_T x_1 x_2 \frac{d\hat{\sigma}}{dt}(a+b\rightarrow{J/\psi}+{J/\psi}).
\label{eqae}
\end{align}
where $y_1$ and $y_2$ are recorded as the rapidity of the produced double $J/\psi$, $f_1(x_1)$ and $f_2(x_2)$ represent parton distribution functions, $p_T$ is a single $J/\psi$ transverse momentum, parton momentum density $x_1 = \sqrt{p_T^2 + m_{J/\psi}^2}(e^{y_1} + e^{y_2})/ \sqrt{S}$ and $x_2 = \sqrt{p_T^2 + m_{J/\psi}^2}(e^{-y_1} + e^{-y_2})/ \sqrt{S}$. $\frac{d\hat{\sigma}}{dt}$ is the differential cross section at parton level~\cite{Brock:1993sz}.

For the outgoing $J/\psi$, one can employ the following projection:
\begin{align}
&v(p_{\bar{c}})\bar{u}(p_c)\rightarrow \frac{1}{2\sqrt{2}}\epsilonsl^*_{J/\psi}(\Psl+2m_c)\times(\frac{1}{\sqrt{m_c}}\psi_{J/\psi} (0))\otimes(\frac{\mathbf{1}_c}{\sqrt{N_c}}).
\end{align}
where the wave function at the origin of $J/\psi$ is $|\psi_{J/\psi}(0)|=\sqrt{|R^{c\bar{c}}(0)|^2/{4\pi}}$, and $\epsilon_{J/\psi}$ is the $J/\psi$ polarization vector with $P\cdot\epsilon = 0$, $P$ is the momentum of $J/\psi$, $\mathbf{1}_c$ is the unit color matrix, we also treat $m_{J/\psi}=2m_c$ approximately.

As mentioned in the introduction, there are some large logarithmic terms in all order perturbation expansion,
and we need to resum them together into a Sudakov factor.
In the rest of this section, we show how to derive the Sudakov factor based on the perturbative QCD.
At the small $q_\perp$ limit,
the $q_\perp$ differential cross sections at LO of $\alpha_s$ expansion can be expressed as follows:
\begin{align}
\frac{d\sigma}{d^2q_\perp dy}&= \sigma_{0} \frac{\alpha_{s} C_A}{2\pi
^2}\int f(x_1) dx_1 f(x_2) dx_2\frac{1}{q^{2}_\perp}[\frac{2(1-\varepsilon_1 + \varepsilon^{2}_1)^2}
{1-\varepsilon_1)_{+}}\delta(1-\varepsilon_2)\nonumber\\
& +\frac{2(1 - \varepsilon_2 + \varepsilon^{2}_2)^2}{(1 - \varepsilon_2)_{+}}\delta(1 - \varepsilon_1) +(2 ln \frac{Q^2}{q^{2}_\perp} )
\delta(1 -\varepsilon_2)\delta(1-\varepsilon_1)],
\end{align}
where $\sigma_0$ is the cross section of the tree level, $\varepsilon_1=Q^2 e^y/x_1\sqrt{S}$, $\varepsilon_2=Q^2 e^{-y}/x_2\sqrt{S}$, and $f(x)$ is parton distribution function. After a fourier transform $W(b,Q^2) = \int d^2 q_\perp e^{-i q_\perp \cdot b_\perp}\frac{d\sigma}{d^2q_\perp dy}$.
\begin{align}
W^{(1)}(b,Q^2)& = \sigma_0\frac{\alpha_s C_A}{\pi}\int dx_1 dx_2 f(x_1) f(x_2)[\varepsilon_1 \zeta(\varepsilon_1)\delta(1-\varepsilon_2)
(-\frac{1}{\epsilon}+ \ln \frac{4e^{-2\gamma E}}{\mu^2 b^2})+(\varepsilon_1 \rightarrow \varepsilon_2)]\nonumber\\
&+\delta(1-\varepsilon_1)\delta(1-\varepsilon_2)[b_0 \ln \frac{b^2 Q^2}{4} e^{2\gamma E} -\frac{1}{2}\ln^2(\frac{Q^2 b^2}{4} e^{2\gamma E}) - \frac{\pi^2}{6}],
\end{align}
where $\mu$ is factorization scale,  $C_A = N_c = 3$, $T_F = \frac{1}{2}$, $b_0 = (\frac{11}{6} C_A - \frac{2}{3}T_F n_f)/N_c$, $n_f$ is the number of quark flavors, $n_f = 5$, $\zeta$ is gluon splitting function.
The scale in $W(b,Q^2)$ can be evolved by the CSS evolution equation~\cite{Collins:1984kg}.
Then we can rewrite $W(b,Q^2)$ as:
\begin{align}
W(b,Q^2) = e^{-S_{sud}(Q^2, b, C_1,\mu^2)} W(b, C_1, \mu^2).
\end{align}
where the Sudakov form factor is
\begin{align}
S_{sud} = \int^{\mu^2}_{C^2_1/b^2} \frac{d\bar{\mu}^2}{\bar{\mu}^2}[\ln(\frac{Q^2}{\bar{\mu}^2}) A (C_1,\bar{\mu}) + B(C_1, \bar{ \mu})].
\end{align}
Here $Q^2$ invariant mass of final $J/\psi$ pair system,  $C_1$ and $C_2$ are two arbitrary parameters, and the range of scale $\bar{\mu}^2$ from $C^2_1/b^2$ to $C^2_2 Q^2$.
The scale $\mu^2$ is resummation scale, in principle it can be arbitrary value, in order to
eliminate the possible large logarithm, we usually set it at the typical scale in the process,
for example here we choose $\mu^2=Q^2$.
 The functions A and B can be calculated via perturbative QCD, $A = \sum\limits_{i = 1}^{\infty} A^{(i)}(\frac{\alpha_s}{\pi})^i$ and $B = \sum\limits_{i = 1}^{\infty} B^{(i)}(\frac{\alpha_s}{\pi})^i$. The $W(b, C_1, \mu^2)$ can be expressed as:
\begin{align}
&W(b, C_1, \mu^2) = \sigma_0 \frac{Q^2}{S}\int\frac{d x}{x}\frac{d x'}{x'}C_{gg} (\frac{x_1}{x}, \frac{b^2}{C_1^2}, C_1, \mu^2) C_{gg} (\frac{x_2}{x'}, \frac{b^2}{C_1^2}, C_1, \mu^2)f(x, \mu') f(x', \mu').
\end{align}

In our work, we choose $C_1 = C_3 = 2 e^{-\gamma E}$, $C_2 = 1$ and $\mu' = C_3/b$ to eliminate additional logarithm terms.
And in our calculation $A^{(1)}$, $A^{(2)}$, $B^{(1)}$ and $C^{(1)}$ are considered:
\begin{align}
&A^{(1)} = C_A,\nonumber\\
& B^{(1)} = -2\beta_0 C_A,\nonumber\\
& C^{(1)}_{gg} = \delta(1 - x),\\
& A^{(2)} = C_A [(\frac{67}{36} - \frac{\pi^2}{12})N_c - \frac{5}{18} N_f].
\end{align}

The differential cross section at small $q_\perp$ region can be expressed as:
\begin{align}
&\frac{d \sigma}{d^2 q_\perp d y} = \frac{1}{(2\pi)^2}\int d^2 b e^{i q_\perp\cdot \vec{b}} W(b, Q^2).
\end{align}
 $W(b, Q^2)$ is derived based on the perturbative QCD. However, when b is large, the perturbation calculation is no longer applicable,
 we need to introduce a nonperturbative function $W^{NP}(b)$, and rewrite $W(b, Q^2)$ as the following equation:
 \begin{align}
&W(b, Q^2) = W(b^*,Q^2)*W^{NP}(b).
\end{align}
Here $b^* = b/\sqrt{1 + (b/b_{max})^2}$ and $b^*$ is always smaller than $b_{max}$.
And we can express the nonperturbative function $W^{NP}(b)$ ~\cite{Su:2014wpa} as:
\begin{align}
&W^{NP}(b) =\exp\left[-\frac{C_A}{C_F}\left(g_{1} b^{2}+g_{2} \ln \left(b / b^{*}\right) \ln \left(Q / Q_{0}\right)+g_{3} b^{2}\left(\left(x_{0} / x_{1}\right)^{\lambda}+\left(x_{0} / x_{2}\right)^{\lambda}\right)\right)\right],
\end{align}
where $g_1=0.21{\rm GeV}^{2}$, $g_2=0.68$, $g_3=-0.12{\rm GeV}^{2}$, $Q^{2}_0 = 2.4 {\rm GeV}^{2}$, $x_{0}=0.0001$ and $\lambda=0.2$~\cite{Su:2014wpa}.
The original $W^{NP}(b)$ in~\cite{Su:2014wpa} is obtained by fitting Drell-Yan data, and we assume that $W^{NP}(b)$ for quark anti-quark annihilation
and gluon fusion processes differs by a factor $\frac{C_A}{C_F}$.
On the other hand, the nonperturbative function $W^{NP}(b)$
only has strong effect in the extremely small $q_\perp$ region around $q_\perp<1$GeV,
in the rest region the dependence on it can be ignored.


\section{NUMERICAL RESULTS }\label{sec:3}

In our work, the software MATHEMATICA, FEYNCALC, FEYNARTS and Cuba-4.2~\cite{Hahn:2004fe} are used.
We use the CTEQ6L1 parton distribution functions~\cite{Lai:1999wy} to simulate initial partons.
Both the renormalization scale and the factorization scale are $\mu =\sqrt{p^2_T + 4m_{J/\psi}^2}$.
The nonperturbative parameter is set as $|R^{c\bar{c}}(0)|^2=0.81$ GeV$^3$~\cite{Eichten:1995ch}, and for charm quark mass we choose $m_c=1.55$ GeV.

For each $J/\psi$ rapidity $y^{J/\psi}$ and transverse momentum $p_T^{J/\psi}$ at the region of $2<y^{J/\psi}<4.5$ and $0 < p_T^{J/\psi} < 10$ GeV at $\sqrt{S}=7$ TeV,
LHCb collaboration measures $\sigma^{J/\psi J/\psi}=5.1 \pm 1.0 \pm 1.1 \mathrm{nb}$~\cite{Aaij:2011yc}.
At the LO of $\alpha_s$ expansion,
we predict $\sigma_{LO}^{J/\psi J/\psi}=4.56$ nb,
which is consistent with the result in Ref.~\cite{Sun:2014gca}.
According to Ref.~\cite{Sun:2014gca}, the next leading order(NLO) correction is about 20\% enhancement comparing to LO contribution.
Therefore, the theoretical prediction agrees with experimental measurement very well.



\begin{figure}[H]
	\centering
	\includegraphics[width=.70\textwidth]{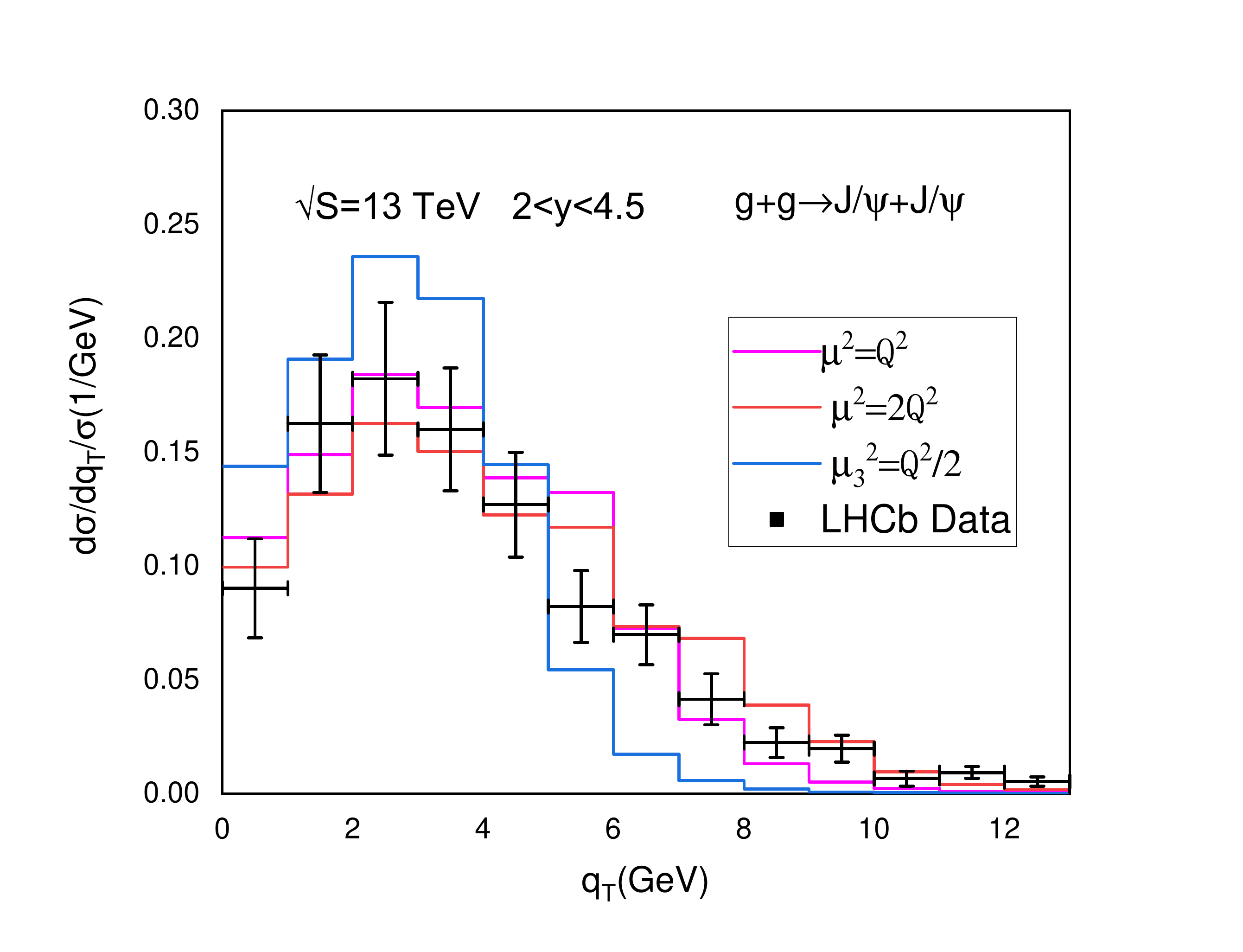}
	\caption{The $q_T$ distribution shape of the double $J/\psi$ in different bins. The data comes from Ref~\cite{Aaij:2016bqq}. The magenta line indicates the resummation result when the resummation scale is $\mu^2=Q^2$. The red and  blue lines are also the resummation results but with the scales  $\mu^2=2Q^2$ and  $\mu^2=\frac{Q^2}{2} $ respectively.  }
	\label{expLHCb}
\end{figure}

The LHCb collaboration also measures the same process at $\sqrt{S}=13$ TeV
with the same cutoff of $y^{J/\psi}$ and $p_T^{J/\psi}$, and
they obtain $\sigma^{J/\psi J/\psi} = 15.2\pm 1.0 \pm 0.9$ nb  ~\cite{Aaij:2016bqq}.
Our prediction is $\sigma_{LO}^{J/\psi J/\psi} = 8.29 $ nb at the LO of $\alpha_s$ expansion.
Comparing to the experimental measurement, our result is smaller by roughly a factor 2.
Although the kinematical cut is as the same as the case at $\sqrt{S}=7$ TeV,
the higher $\sqrt{S}$ could lead to larger high order $a_s$ correction.
And LHCb also measure the $q_T$ distribution at $\sqrt{S}=13$ TeV.
The shape of $q_T$ distribution is decided by Sudakov factor,
and our prediction is consistent with the experimental result as shown in Figure~\ref{expLHCb}.

\begin{figure}[H]
	\centering
	\includegraphics[width=.70\textwidth]{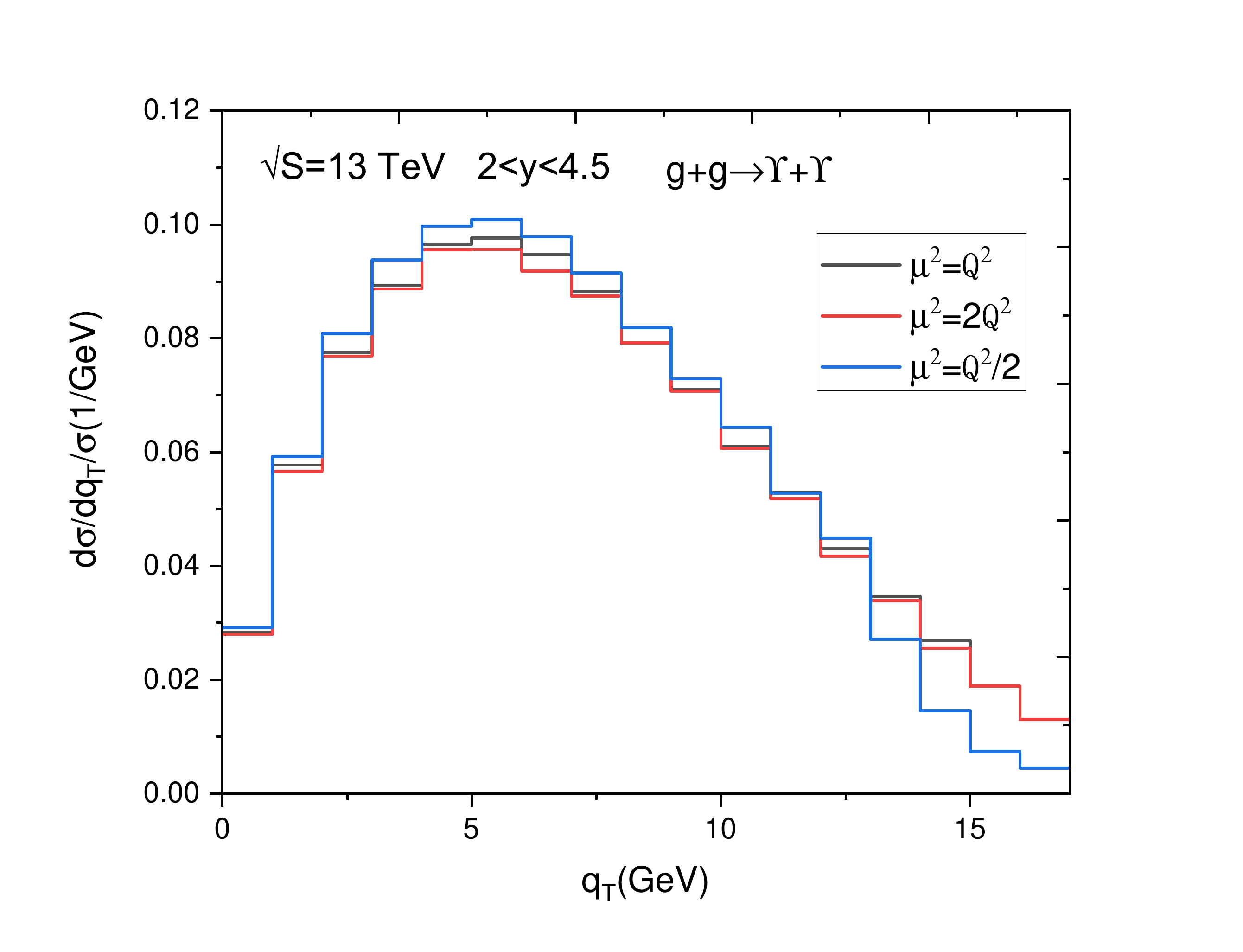}
	\caption{The $q_T$ distribution shape of the double $\Upsilon$ in different bins. The black line indicates the resummation result when the rsummation scale is $\mu^2=Q^2$. The red and  blue lines are also the resummation results but with the resummation scales $\mu^2=2Q^2$ and  $\mu^2=\frac{Q^2}{2}$ respectively.}
	\label{uplson}
\end{figure}
The CMS collaboration measures the cross section production of double $\Upsilon$ at $\sqrt{S}=13$ TeV  with the cutoff
$ \mid y^{\Upsilon(1S)}\mid < 2.0 $, $0 < p^{\Upsilon(1S)}_T < 10$ GeV, and they obtain
$\sigma_{\mathrm{fid}}=79 \pm 11(\text { stat }) \pm 6(\text { syst }) \pm 3(\mathcal{B}) \mathrm{pb}$
~\cite{Sirunyan:2020txn}.
 With $m_b=4.6$ GeV, $|R^{b\bar{b}}(0)|^2=6.477$ GeV$^3$~\cite{Eichten:1995ch},
 we obtain the total cross section $\sigma_{LO}^{\Upsilon \Upsilon} = 54.9$ pb at the LO of $\alpha_s$ expansion.
 We calculate the $q_\perp$ distribution using the method of resummation.
 In Figure ~\ref{uplson}, we draw the $q_\perp$ distribution shape of double $\Upsilon$ production at LHC.


\section{Summary and Conclusions} \label{sec:6}

In the framework of NRQCD theory, we study the production of double heavy quarkonia at LHC.
Using the resummation technique, the $q_\perp$ distribution of double heavy quarkonia pair system
is calculated. At $\sqrt{S}= 13$ TeV with $0< p_T^{J/\psi} < 10$ GeV and $2<y^{J/\psi}<4.5$,
the resummation result can simulate experimental data of $q_\perp$
distribution shape very well. 
Since we only consider each $J/\psi$ with small transverse momentum $0< p_T^{J/\psi} < 10$ GeV,
the color octet effect can be ignored here, and we only consider the contribution that $J/\psi$
comes from color-singlet charm quark pair.
In this work, except large logarithm terms generated by soft gluon radiation,
the higher order $\alpha_s$ correction is not included.
 Such contribution could be important
in large $q_\perp$ region. However in small $q_\perp$ region the
distribution shape is mainly determined by the Sudakov factor.
On the other hand, in order to give a precise prediction of $q_\perp$ distribution,
we do need to take into account the higher order $\alpha_s$ correction and color
octet contribution. We will study them in our future work. Through this 
study we know the QCD resummation can describe the soft gluon radiation for the gluon fusion 
process very well, it supply important support for us to use the same method 
to study the other gluon fusion process like Higgs production at LHC.



\begin{acknowledgments}

P. Sun is supported by Natural Science Foundation of China under grant No. 11975127 as well as Jiangsu Specially Appointed Professor Program.
This work is also supported by NSFC under grant No.~12075124. and by Jiangsu Qing Lan Project.

\end{acknowledgments}


\end{document}